%% file: Manuscript.tex
\documentclass[12pt]{article}

\RequirePackage[OT1]{fontenc}
\usepackage{natbib}
\RequirePackage[colorlinks,citecolor=blue,urlcolor=blue]{hyperref}
\usepackage{xurl}
\usepackage[english]{babel}
\usepackage{amsthm}
\usepackage{amsmath}
\usepackage{amsfonts}
\usepackage{amssymb}
\usepackage{graphicx}
\usepackage{enumerate}
\usepackage{color}
\usepackage{array}
\usepackage{tikz}
\usepackage{comment}
\usepackage{bm}
\usepackage{multirow}
\usepackage{booktabs} 
\usepackage{adjustbox}
\usepackage{titlesec}
\usepackage{subcaption}
\usepackage{caption}
\usepackage{float}

\definecolor{myorange}{HTML}{E24100}
\definecolor{mygreen}{HTML}{228b22}

\input{macros.tex}

\input{teachingmacros.tex}

\usepackage[margin=1in]{geometry}

\begin{document}
\thispagestyle{empty}
\baselineskip=26pt
\vskip 5mm
\begin{center} 
{\Large{\bf{Spatial modeling and future projection of extreme precipitation extents}}}
\end{center}

\baselineskip=12pt
\vskip 5mm

\begin{center}
\large
Peng Zhong$^1$, Manuela Brunner$^{2, 3}$, Thomas Opitz$^4$, and Rapha\"el Huser$^1$ 
\end{center}

\footnotetext[1]{
\baselineskip=10pt Statistics Program, Computer, Electrical and Mathematical Sciences and Engineering (CEMSE) Division, King Abdullah University of Science and Technology (KAUST), Thuwal 23955-6900, Saudi Arabia. E-mails: peng.zhong@kaust.edu.sa; raphael.huser@kaust.edu.sa}
\footnotetext[2]{
\baselineskip=10pt Institute for Atmospheric and Climate Science, ETH Zurich, Zurich, Switzerland, E-mail: manuela.brunner@env.ethz.ch}
\footnotetext[3]{
\baselineskip=10pt Institute for Snow and Avalanche Research SLF, Swiss Federal Institute for Forest, Snow and Landscape Research WSL, Davos, Switzerland}
\footnotetext[4]{
\baselineskip=10pt Biostatistics and Spatial Processes, INRAE, Avignon, 84914, France, E-mail: thomas.opitz@inrae.fr}

\baselineskip=16pt
\vskip 4mm
\centerline{\today}
\vskip 6mm

\begin{center}
{\large{\bf Abstract}}
\end{center}

Extreme precipitation events with large spatial extents may have more severe impacts than localized events as they can lead to widespread flooding. It is debated how climate change may affect the spatial extent of precipitation extremes, whose investigation often directly relies on simulations from climate models. Here, we use a different strategy to investigate how future changes in spatial extents of precipitation extremes differ across climate zones and seasons in two river basins (Danube and Mississippi). We rely on observed precipitation extremes while exploiting a physics-based mean temperature covariate, which enables us to project future precipitation extents. We include the covariate into newly developed time-varying $r$-Pareto processes using a suitably chosen spatial aggregation functional $r$. This model captures temporal non-stationarity in the spatial dependence structure of precipitation extremes by linking it to the temperature covariate, which we derive from observations for model calibration and from debiased climate simulations (CMIP6) for projections. For both river basins, our results show negative correlation between the spatial extent and the temperature covariate for most of the rain season and an increasing trend in the margins, indicating a decrease in spatial precipitation extent in a warming climate during rain seasons as precipitation intensity increases locally.

\baselineskip=16pt
\par\vfill\noindent

\pagenumbering{arabic}
\baselineskip=26pt

\newpage

\allowdisplaybreaks
\baselineskip=26pt
\clearpage
\section{Introduction}\label{sec:intro}
Extreme precipitation events with large spatial extents may have more widespread and severe impacts than localized events, which is why they are associated with greater management challenges. For example, they may lead to widespread flooding requiring the coordination of evacuation measures across river basins. Increases in the frequency and magnitude of extreme precipitation events are evident both in observations \citep{Contractor.etal:2021, Kirchmeier.Zhang:2020, Myhre.etal:2019, Zeder.Fischer:2020, Papalexiou.Montanari:2019} and future model simulations \citep{Bao.etal:2017, Prein.etal:2017, Wood.Ludwig:2020, Brunner.etal:2021, Pendergrass.etal:2019, Swain.etal:2018,Na.etal:2020}. While precipitation intensities and the frequency of extreme events have been shown to increase in wide parts of the world, it remains less clear how the spatial extent of these events will change in a warming climate. Several observation-based and model-based studies have suggested that the spatial extent of extreme precipitation events changes as a result of warming temperatures. However, the direction of this change is yet unclear.
\citet{Wasko.etal:2016} have shown that observed precipitation extents in Australia decrease with temperature, while \citet{Tan.etal:2021} have demonstrated increases in the observed spatial extent of precipitation extremes for several regions over the Northern Hemisphere and in the western Pacific over 1983--2018. Such increases have also been found by \citet{Lochbihler.etal:2017} who have revealed a clear relationship between event intensity and spatial extent using radar data over the Netherlands. Discrepancies in changes of precipitation extents are not limited to observation-based studies but extend to model-based studies predicting the potential future evolution of precipitation spatial extents.
\citet{Chang.etal:2016} have modeled a decrease in storm size over large parts of North America under climate change and \citet{Guinard.etal:2015} have projected both decreases and increases in storm area for different regions in North America. In contrast, \citet{Bevacqua.etal:2021} have projected increases in the spatial extent of wintertime precipitation extremes over the Northern Hemisphere. These contrasting observed trends and future projections of spatial extents of extreme precipitation may be a result of different event and extent definitions \citep{Rastogi.etal:2020}. In addition, they may result from a focus on different seasons and regions as spatial extents of extreme precipitation events have been shown to vary both seasonally and regionally \citep{Chang.etal:2016, Touma.etal:2018, Rastogi.etal:2020, Tan.etal:2021}. Even though season and region may be important determinants of future changes in spatial extents of extreme precipitation, most existing studies focus on particular regions and do not differentiate between different seasons. Therefore, we here explore how future changes in spatial extents of precipitation extremes differ for different climate zones and seasons. 

Past studies projecting future changes in extreme precipitation extents have mainly focused on climate model outputs for precipitation, even though regional climate models may substantially  underestimate or overestimate the spatial dependence of extremes depending on the season \citep{Yang.etal:2020}. 
Here, we use a different strategy based on extreme-value theory to study how future changes in spatial extents of precipitation extremes differ across climate zones and seasons. We rely on observed precipitation extremes and exploit a physics-based temperature covariate derived from climate model output to obtain future projections of spatial precipitation extents, as the simulation of temperature variables by climate models is generally considered as more reliable than the simulation of precipitation extremes
\citep{Aloysius.elal:2016,Stephens.etal:2010}. While extreme-value theory has been frequently used to investigate temporal trends in extreme rainfall events \citep[e.g.,][]{Helga.Holger.David:2021}, the focus has been on detecting trends in the margins. Here, we focus on changes in the spatial characteristics of precipitation when marginal trends have already been accounted for. We associate larger extents with stronger spatial correlation among extreme values, and therefore with a longer tail-correlation range. If a relatively large precipitation intensity occurs at a given location, other locations with a high positive tail-correlation to the given location will also tend to show relatively large precipitation intensities. Therefore, if the tail-correlation range is longer, large intensities will also occur at locations that are relatively far from the given location, leading to a larger spatial extent of the overall precipitation event. To get an objective measure of the extent of spatial extreme events, we thus propose to compute the \emph{effective tail-correlation range}, defined as the minimum distance at which the tail-correlation drops below $0.05$. In this paper, we concentrate on precipitation extents in two river basins from different continents and climate zones, namely the Danube basin (temperate-humid climate, Europe) and the Mississippi basin (continental climate transitioning to humid subtropical, North America) as changes in precipitation extents may at least partly lead to changes in widespread flooding \citep{Brunner.etal:2020}, and we estimate the effective tail-correlation range in each basin--season case separately.

There are two main ways to model spatial extremes in the literature; one uses pointwise block maxima \citep{Davison.etal:2012,Davison.Huser:2015,Davison.etal:2019, Huser.etal:2022} and the other one uses exceedances over a high threshold \citep{Davison.Smith:1990,Huser.Davison:2014a,Opitz2015,Thibaud.Opitz:2015,Richards.etal:2022}. Computing pointwise block maxima from a complex dataset can lead to a significant loss of information, which can significantly undermine the effort to detect trends in the dependence structure. In particular, it may not be easy to identify suitable covariates in a regression context when daily data are aggregated (through the maximum operator) to a monthly, seasonal, or yearly scale. Moreover, computing pointwise block maxima also requires a relatively complete dataset. Selecting a complete sub-dataset can result in a significant loss of information if the dataset contains many missing values. Therefore, modeling exceedances over a high threshold is a preferable approach in this non-stationary trend detection context as it prevents such loss of information. In this work, we develop a method based on $r$-Pareto processes \citep{Ferreira.deHaan:2014, Dombry.Ribatet:2015} to model extreme precipitation peaks-over-threshold and their spatial characteristics. \cite{deFondeville.Davison:2018} developed a fast score-matching inference method for a class of $r$-Pareto processes associated with log-Gaussian random functions that can be applied in high spatial dimensions. Here, we extend this approach to incorporate a time-varying semivariogram in the dependence structure. Such a semivariogram allows us to estimate the time-varying spatial extent of precipitation extremes by incorporating a well-chosen aggregated temperature covariate into the dependence model. We apply this model to predict the spatial extent of precipitation extremes under different climate change scenarios. Specifically, to derive future projections of precipitation extents, we consider historical and future climate model runs (using different CMIP6 simulations) based on the Shared Socioeconomic Pathways (SSP) 2-4.5 and 5-8.5, which represent ``middle-of-the-road'' and more pessimistic ``fossil-fueled development'' scenarios, respectively, and we then report projected changes in the effective tail-correlation range under each scenario. 

This paper is organized as follows: we present the dataset and application examples in Section~\ref{sec:Dataset}. In Section~\ref{sec:method}, we first detail how our physics-based temporal covariate is designed, and we then describe the non-stationary marginal and $r$-Pareto dependence models fitted to our precipitation dataset. For margins, we adopt a three-step generalized additive modeling strategy and detail each of the steps precisely. In Section~\ref{sec:results}, we present the results from the dependence model fit and report the estimated spatial extent of precipitation extreme events under different climate change scenarios for the different basin--season cases under study. We discuss statistical and hydro-meteorological considerations in Section~\ref{sec:discussion}, and we finally conclude in Section~\ref{sec:conclude} with some perspective on future research and possible extensions.

\section{Dataset}\label{sec:Dataset}
\subsection{General description}\label{subsec:data_description}
\begin{figure}[!t]
            \centering
            \begin{subfigure}{0.4\linewidth}
                \hspace{-0.6cm}\includegraphics[height=0.25\textheight]{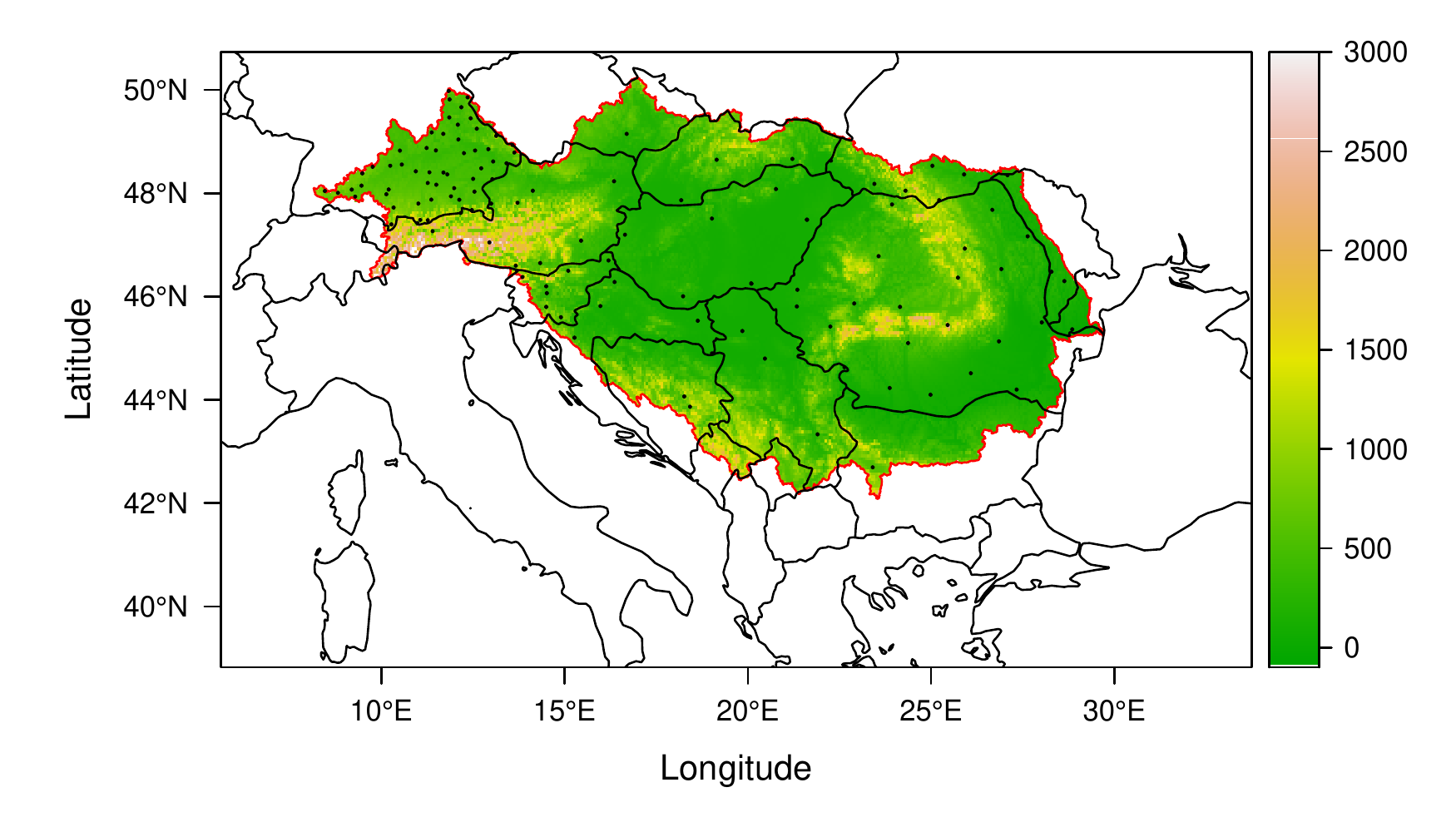}
                \caption{Map of Danube river basin.}
                \label{fig:Danube}
            \end{subfigure}
            \hfill
            \begin{subfigure}{0.4\linewidth}
                \hspace{-2cm}\includegraphics[height=0.25\textheight]{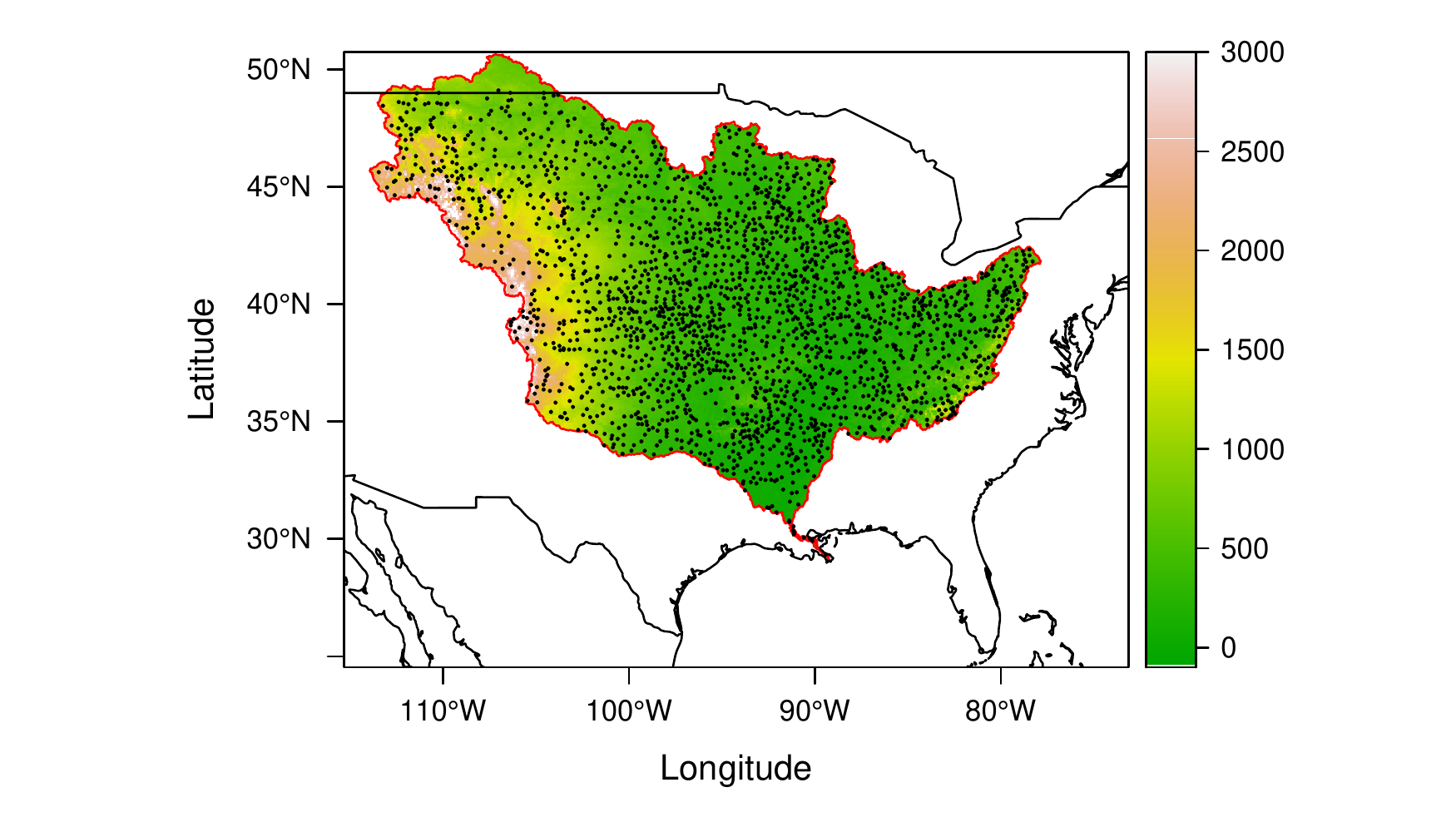}
                \caption{Map of Mississippi river basin.}
                \label{fig:Mississippi}
            \end{subfigure}
            \caption{Map of Danube (left) and Mississippi (right) river basins with black dots representing the precipitation gauges. The color scale indicates elevation (m.a.s.l.).}
            \label{fig:basins}
\end{figure}

We focus on two large river basins in different climate zones whose characteristics are summarized in Table~\ref{tab:basins} and Figure~\ref{fig:basins}. The dataset consists of observed daily precipitation data in millimeters and daily temperature averages in degrees Celsius from 125 monitoring stations in the Danube river basin (Europe) and from 2229 monitoring stations in the Mississippi river basin (North America). The dataset is publicly available from the \href{https://www.ncdc.noaa.gov/ghcnd-data-access}{Global Historical Climatology Network (GHCN)} for the period 1965--2015. There are in total $60\%$ of missing values in the Danube region and $77\%$ of missing values in the Mississippi region, which still leaves a considerable number of non-missing values during the whole observation period given the length of the time series and the number of monitoring stations. 
\begin{table}[!t] 
 \caption{Summary of characteristics of the two study regions: region name, continent, climate zone, area (km$^2$), average elevation given as meters above sea level (m.a.s.l.), and the number of available precipitation and temperature gauges.}
\begin{adjustbox}{width=\textwidth}
\begin{tabular}{ llllll } 
 \hline
 \textbf{Region} & \textbf{Continent} & \textbf{Climate zone} & \textbf{Area (km$^2$)} & \textbf{Elevation (m.a.s.l.)} & \textbf{Number of gauges} \\ 
 \hline
 \textbf{Danube} & Europe & Temperate-humid & 797335 & 462 & 125\\ 
 \textbf{Mississippi} & North America & Continental to subtropical & 3244506 & 682 & 2229 \\
 \hline
\end{tabular}
\end{adjustbox}
 \label{tab:basins} 
\end{table}

\subsection{Exploratory analysis}\label{subsec:explore}
\begin{figure}[!t]
            \centering
          \includegraphics[width=0.49\linewidth,page=2]{heatmaps.pdf}
            \includegraphics[width=0.49\linewidth,page=4]{heatmaps.pdf}
            \includegraphics[width=0.49\linewidth,page=1]{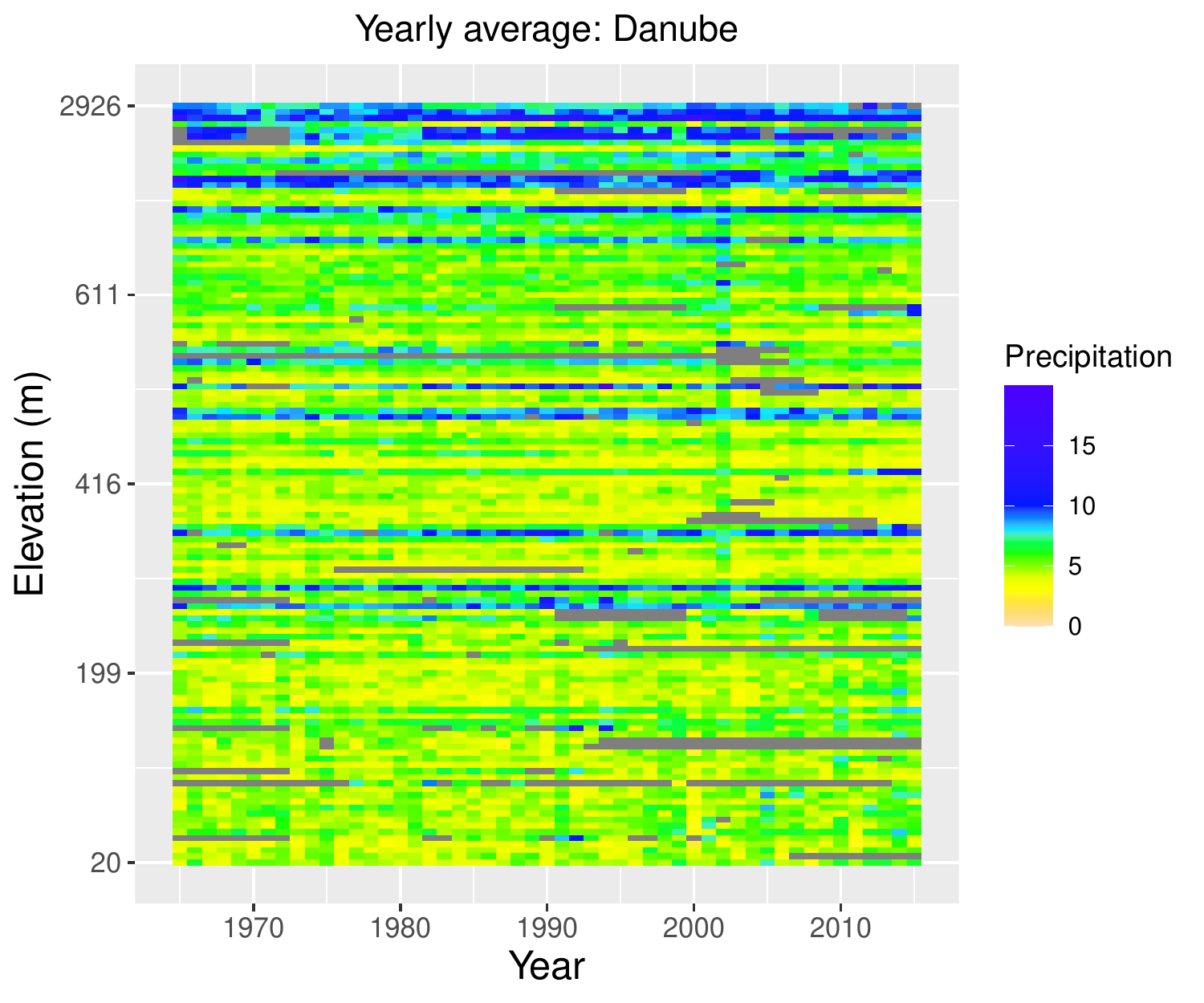}
            \includegraphics[width=0.49\linewidth,page=3]{heatmaps.pdf}
            \caption{Heatmaps illustrating the seasonal (top) and year-to-year (bottom) variations of daily precipitation intensities (mm) for the Danube basin (left) and Mississippi basin (right). Top: values averaged across years, i.e., $\bar Y_{i,\cdot,k}=\frac{1}{51}\sum_{j=1}^{51}Y_{i,j,k}$, plotted for each day $i$ of the year (x-axis), and station $\bm s_k$ sorted by elevation (y-axis). Bottom: values averaged over each day of the year, i.e., $\bar Y_{\cdot,j,k}={1\over n_j}\sum_{i=1}^{n_j}Y_{i,j,k}$ with $n_j\in\{365,366\}$, plotted for each year $j$ (x-axis), and station $\bm s_k$ sorted by elevation (y-axis). Blue, green, yellow, and grey colors represent high, medium, low, and missing precipitation values, respectively.}
            \label{fig:summary}
\end{figure}

Before developing a model for spatial precipitation, we explore the characteristics of daily precipitation in the two river basins, i.e., temporal trends, seasonal patterns, and possible correlations with temperature. Let $\calS\subset\Real^2$ denote the spatial domain under study (either the Danube basin or the Mississippi basin), and $\calD=\{\bm s_1,\ldots,\bm s_K\}\subset\calS$ be the set of monitoring stations, with $K=125$ (Danube basin) and $K=2229$ (Mississippi basin). We write $Y_{i,j,k}$ to denote the daily precipitation amount on the $i$-th day during the $j$-th year at the $k$-th station $\bm s_k$, where $j\in \{1,\dots,51\}$, $i\in \{1,\dots,n_j\}$ with $n_j$ either equal to 365 or to 366 for  leap years,  and $k\in\{1,\ldots,K\}$. In order to conveniently visualize the seasonal behavior of the precipitation data across all stations in a single plot, we compute the daily average $\bar Y_{i,\cdot,k} = \frac{1}{51} \sum_{j=1}^{51}Y_{i,j,k}$ for each day $i$ and station $\bm s_k$, and then show these values as heatmaps in the top panels of Figure~\ref{fig:summary}. This helps to explore the presence of any seasonal patterns in the data. 
Due to the high percentages of missing values in the dataset, we only compute the average when at least 10 data points are available. Otherwise, we treat the average $\bar Y_{i,\cdot,k}$ as missing. Precipitation displays a clear seasonal pattern and variation with respect to the elevation of the monitoring stations. Both river basins show the strongest precipitation intensities during the summer. The Danube river basin is characterized by slightly increasing precipitation with increasing elevation (i.e., lower temperature). In contrast, the Mississippi river basin shows increasing precipitation with decreasing elevation (i.e., higher temperature). We also compute the annual average $\bar Y_{\cdot,j,k} = {1\over n_j} \sum_{i=1}^{n_j}Y_{i,j,k}$ for each year $j$ and station $\bm s_k$ to explore whether there is any global temporal trend in precipitation intensities. We only calculate the average if at least 20 data points are available in this case. The bottom panels of Figure~\ref{fig:summary} show heatmaps of $\bar Y_{\cdot,j,k}$, revealing patterns associated with elevation that are consistent with those identified in the top panels of Figure~\ref{fig:summary}. An annual temporal trend signal, however, cannot be clearly detected from these visual diagnostics. Thus, potential temporal trends in the margins and the dependence structure of extreme precipitation remain to be assessed using more sophisticated extreme-value regression models. Our approach is discussed in the following sections.

\section{Methodology}\label{sec:method}

\subsection{Designing a suitable physics-based temporal covariate}\label{sec:covariate}
\begin{figure}
    \centering
    \includegraphics[page=4,width=0.48\linewidth]{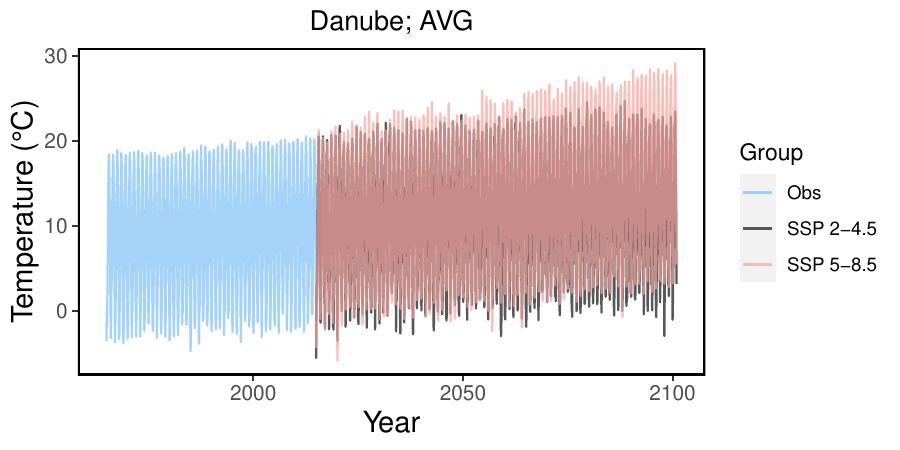}
    \includegraphics[page=8,width=0.48\linewidth]{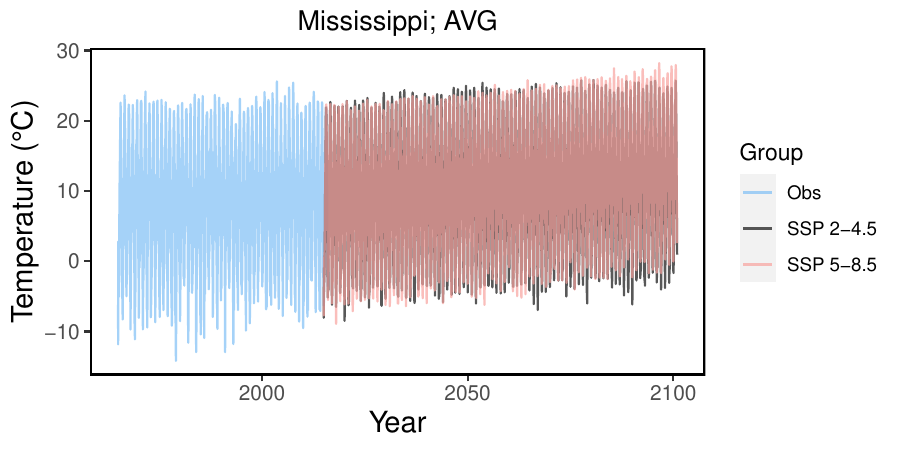}
    \caption{Plots of the temporal covariate, $\text{temp}_t$, for the Danube river basin (left) and Mississippi river basin (right) based on actual temperature observations (light blue), as well as the projected temperature covariate between the year of 2016 and the year of 2100, averaged across simulations of three different CMIP6 climate models runs, namely AWI, MIROC, and NorESM, under the Shared Socioeconomic Pathways (SSP) 2-4.5 (dark gray) and 5-8.5 (light red), described in detail in  Section~\ref{subsec:futureprojections}.}
    \label{fig:temp_covariate}
\end{figure}

The choice of a temporal covariate for use in marginal (Section~\ref{subsec:marginal_model}) and dependence (Section~\ref{subsec:dep_model}) modeling is crucial because it determines the form of nonstationarity that the model can capture. In addition, it drives future projections, thus impacting our conclusions about the evolution of precipitation intensities and spatial extents. Hence, we need to carefully design a covariate that (i) has a physical meaning; (ii) is relevant for predicting extreme precipitation intensities and extents (i.e., it must be ``correlated'' to intensities and spatial extents, both expressed on the logarithmic scale here); (iii) reflects climate conditions across the whole river basin under study; and (iv) can be relatively easily projected into the future in a physically justifiable way under various climate change scenarios. Spatially-aggregated, basin-specific air temperature satisfies all of these four requirements, as several studies have highlighted the physical link between temperature and precipitation amounts, which is also confirmed in Figure~\ref{fig:summary}. There is indeed a wide consensus that global warming will lead to an increase in the water holding capacity of the atmosphere, and thus to more intense precipitation \citep{Pendergrass.etal:2018, IPCC:2021, Fowler.etal:2021, Muller.etal:2011}. The regional response of precipitation to global warming may vary, but a positive correlation between temperature and precipitation intensity can in general be expected, even though the link with spatial extents is less clear. Moreover, air temperature is among the variables that can be the most reliably reproduced and predicted with climate models, though there are still often systematic biases \citep{Hausfather.etal:2020}. We note, however, that affine time-independent biases corresponding to systematic shift or rescaling of the ``true'' values, e.g., when modifying the temporal covariate $\text{temp}_t$ as $a+b\times \text{temp}_t$ for some constants $a\in\Real$ and $b\not=0$, would still produce the same extrapolation if combined with a generalized linear model of the form $\eta=g(\lambda_0+\lambda_1\times\text{temp}_t)$ for some link function $g(\cdot)$, as the biases would simply be absorbed into the intercept and slope coefficients $\lambda_0$ and $\lambda_1$ but the resulting estimate of $\eta$ would remain unchanged. Nevertheless, biases are typically not perfectly time-independent and may also vary spatially (though this might cancel out after spatial aggregation). Furthermore, temperatures simulated from historical climate model runs may be able to represent long-term trends, but often lack correspondence with the actual observations, because climate models not conditioned on observational weather data are ``climate simulators'' rather than ``weather simulators''. These issues are problematic for detecting a meaningful association between observed precipitation and simulated temperature. Therefore, temperature data from climate model outputs may not be the most suitable choice for modeling observed precipitation extremes in the historical period (i.e., for model fitting), while still being very helpful (after adjustment) for future extrapolation. Hence, we here choose to fit our model (further detailed in Sections~\ref{subsec:marginal_model} and \ref{subsec:dep_model}) using a temperature covariate derived from real daily measurements at the same monitoring sites as our precipitation data, but we then use (properly debiased) temperatures from climate model outputs (under different greenhouse gas emission scenarios) for future extrapolation. More details on the choice of climate models and climate change scenarios used for future projections, as well as our simple bias-correction procedure, are provided in Section~\ref{sec:results}. In general, climate model outputs show less variability than observations at weather stations, so we here use spatial temperature averages over the entire river basin, such that the variability of the temperature covariate can be assumed to be comparable between the two data sources. To compute basin-wide temperature averages, a practical problem is that observed daily temperatures are not available at every location within the spatial domain and time point during the historical period, and they also contain almost the same number of missing values as the precipitation data. To overcome these issues, we used a kriging scheme to impute missing temperature observations spatially, for each time point separately. Specifically, we first fit a spatial generalized additive model (GAM) with a Gaussian response distribution and identity link function, defined as
\begin{equation}\label{eq:kriging}
	\text{temp}_i = f(\text{lon}_i,\text{lat}_i) + f(\text{elev}_i) + f(\text{day}_i) + f(\text{year}_i) + \varepsilon_i, \qquad i=1,2,\ldots,
\end{equation}
where $\text{temp}_i$ is the $i$-th temperature measurement, characterized by its longitude ($\text{lon}_i$), latitude ($\text{lat}_i$), and elevation ($\text{elev}_i$) of the corresponding monitoring station, as well as the time of observation (calendar day, $\text{day}_i$, and year, $\text{year}_i$), where $f$ represents penalized cubic spline functions (with bivariate splines implemented as tensor products), and $\varepsilon_i$ denotes independent and identically distributed (i.i.d.) zero-mean Gaussian noise. After this model is estimated, we fit a (stationary) spatial exponential covariance function to the fitted residuals $\hat\varepsilon_i$, treating the spatial replicates as independent. Finally, we use this Gaussian model to interpolate the mean temperature on a fine grid within each river basin for each day separately conditional on the observed data. Specifically, we apply our kriging scheme and interpolate daily temperatures at 442 and 1625 gridded locations in the Danube and Mississippi regions, respectively, which represent a resolution of $0.4622^\circ\times0.4622^\circ$ approximately in latitude and longitude. Then, we use these spatially-imputed daily temperatures to estimate basin-wide temperature averages for each day. Finally, we average for each day $t$, the kriged basin-wide temperature over the 30-day time window preceding the day of the event (i.e., ending at time index $t$) to account for the fact that extreme precipitation events may be the result of several consecutive days of ``favorable'' climate conditions. Our final temporal covariate as shown in Figure~\ref{fig:temp_covariate}, denoted by $\text{temp}_t$, is then the spatiotemporal average of daily temperature over the entire basin and a monthly moving window, further standardized by subtracting its mean and dividing by its standard deviation to stabilize inference. We consider this 30-day moving window a good indicator of the climate conditions in each river basin at each time $t$ and therefore expect this temperature covariate to be correlated with extreme precipitation and to provide a sound framework for future extrapolation. Figure~\ref{fig:temp_covariate} also shows the projected temperature covariate until the end of the 21st century, averaged across simulations of three different climate model runs under the Shared Socioeconomic Pathways (SSPs) 2-4.5 and 5-8.5. Temperature projections for each individual climate model run are shown in Figure~1 of the Supplementary Material. This plot indicates that climate change seems to affect the Danube river basin more strongly than the Mississippi river basin. 

\subsection{Marginal modeling}\label{subsec:marginal_model}
We now detail our marginal model, before presenting how spatial dependence is modeled in Section~\ref{subsec:dep_model}. As shown in Figure~\ref{fig:summary}, the precipitation data show a clear seasonality and are related to elevation. Hence, it is important to account for the non-stationarity in the marginal distributions. Our proposed marginal model is similar to the three-step model of \cite{Opitz.etal:2018}, except that we use a semi-parametric frequentist approach rather than a Bayesian latent Gaussian model. Let $Y_{\bm s,t}$ denote the precipitation data (mm) at time $t$ and site $\bm s\in\calS$, where $\calS$ denotes the spatial domain. Moreover, through our modeling, we found that more than 80\% of the precipitation data are smaller than 10mm. By removing the data that are below 10mm, the marginal fits are considerably improved, and also makes the computational burden much lighter---we thus initially ignore these values. Our overall marginal model may be built in consecutive steps using the following three GAMs, defined in terms of different response distributions based on the underlying linear predictors $\eta_{\bm s,t}^{\text{Gam}}$, $\eta_{\bm s,t}^{\text{Log}}$ and $\eta_{\bm s,t}^{\text{GP}}$, respectively, capturing spatiotemporal characteristics in the bulk and the tail:
	\begin{enumerate}[(i)]\label{eq:margin_model}
		\item \textbf{Gamma model for the bulk:} we assume that $(Y_{\bm s,t} - 10)  \mid Y_{\bm s,t}>10  \sim \text{Gamma}$, with spatiotemporal mean $\exp(\eta_{\bm s,t}^{\text{Gam}})$ and constant shape parameter $\kappa>0$. We have found that the Gamma distribution performs quite well in our case, especially given that it here only serves the purpose of estimating a high spatiotemporal quantile $u_{\bm s,t}$. 
		\item \textbf{Logistic model for occurrence indicators of high threshold exceedances:} we assume that $\mathbb I(Y_{\bm s,t} > u_{\bm s,t}) \sim \text{Bernoulli}$, with spatiotemporal mean $\text{logit}^{-1}(\eta_{\bm s,t}^{\text{Log}})=\exp(\eta_{\bm s,t}^{\text{Log}})/\{1+\exp(\eta_{\bm s,t}^{\text{Log}})\}$, where $u_{\bm s,t}$ is the estimated 90\% quantile from the Gamma model detailed in (i), and $\mathbb I(\cdot)$ is the indicator function;
		\item \textbf{Generalized Pareto (GP) model for high threshold exceedances:} we assume that $(Y_{\bm s,t} - u_{\bm s,t}) \mid Y_{\bm s,t} > u_{\bm s,t} \sim \text{GP}$,
		with spatiotemporal scale parameter $u_{\bm s,t}\exp(\eta_{\bm s,t}^{\text{GP}})$ and constant shape parameter $\xi\in\Real$; that is, $\Pr(Y_{\bm s,t}> u_{\bm s,t}+y \mid Y_{\bm s,t} > u_{\bm s,t})=[1+\xi y/\{u_{\bm s,t}\exp(\eta_{\bm s,t}^{\text{GP}})\}]^{-1/\xi}$, for $0<y<y_+$, with $y_+=\infty$ if $\xi\geq0$ and $y_+=-u_{\bm s,t}\exp(\eta_{\bm s,t}^{\text{GP}})/\xi$ if $\xi<0$, and $u_{\bm s,t}$ defined as in (ii). 
		
		We here use the GP distribution because it is supported by extreme-value theory as the only possible limiting distribution for (properly rescaled) high threshold exceedances, when the threshold tends to the upper endpoint of the distribution. Therefore, it guarantees robust marginal tail extrapolations.
	\end{enumerate}
In the model specifications above, each of the terms $\eta_{\bm s,t}^{\text{Gam}}$, $\eta_{\bm s,t}^{\text{Log}}$ and $\eta_{\bm s,t}^{\text{GP}}$ is assumed to follow the canonical form
$$\eta_{\bm s,t}=f(\text{lon}_{\bm s}, \text{lat}_{\bm s}) + f(\text{elev}_{\bm s}) + f(\text{day}_t) + \beta\times\text{temp}_{t},$$
adopting a notation similar to \eqref{eq:kriging}, but using a slightly different structure than in the kriging temperature model. In particular, we here include the temperature covariate, $\text{temp}_t$, as a linear fixed effect with regression coefficient $\beta$, in place of the nonlinear yearly effect, given that $\text{temp}_t$ can capture time trends that we can then more easily project into the future using by climate models. Such GAMs can be conveniently and efficiently fitted to data observed at the monitoring sites $\calD=\{\bm s_1,\ldots,\bm s_K\}\subset\calS$ using standard functions in the \texttt{R} package \texttt{evgam}. As demonstrated in Section~\ref{subsec:marginalestimation}, the proposed marginal model fits the data very well and captures their spatiotemporal characteristics satisfactorily. 

\subsection{Dependence modeling with time-varying $r$-Pareto processes}\label{subsec:dep_model}
We now model the spatial dependence structure of extreme precipitation, in order to estimate their spatial precipitation extent and assess whether it has changed over time. To this end, we model spatial threshold exceedances defined in terms of a ``risk functional'' $r$ using $r$-Pareto processes associated with log-Gaussian stochastic processes \citep{Dombry.Ribatet:2015, deFondeville.Davison:2018}, whose max-stable counterparts are the so-called Brown--Resnick processes \citep{Brown.Resnick:1977,Kabluchko:2009}. Modeling $r$-threshold exceedances allows us to keep more information and to borrow more strength across the spatiotemporal domain in comparison with the classical block maximum approach, while also keeping flexibility in the way spatial extreme events are defined through $r$. The risk functional $r$ must be nonnegative and homogeneous but is otherwise arbitrary, and can, for example, be the spatial average, maximum, or minimum over the entire domain $\calS$ or a (potentially finite) subdomain, such as $\calD=\{\bm s_1,\ldots,\bm s_K\}\subset\calS$ corresponding to the monitoring stations themselves. The theoretical justification for using $r$-Pareto processes is that they naturally appear as the only possible limits of (renormalized) spatially-indexed threshold exceedances as the threshold increases arbitrarily. This result directly extends the univariate GP limit distribution to the spatial setting, and we summarize it in the following theorem. 
We denote by $\calC_+(\calS)$ the space of continuous nonnegative functions on $\calS$, and let $r$ be a (nonnegative and homogeneous) risk functional on $\calC_+(\calS)$. 
\begin{theorem}[\cite{Dombry.Ribatet:2015}]\label{thm:rPareto}
Let $Y$ be a random process defined on a compact nonempty domain $\calS$ with $\alpha$-Pareto margins, i.e., $\Pr(Y>y)=y^{-\alpha}$, $y\geq1$, for some $\alpha>0$, and let $r$ be a continuous and homogeneous risk functional. If
$$\Pr\left(u^{-1} Y\in \cdot \mid r(Y) > u\right) \to \Pr \left(Z \in \cdot\ \right),\ u\to\infty,$$
with weak convergence in  $\mathcal \calC_+(\calS)$, then, either $\Pr\left(r(Z) = 1\right) = 1$ or $Z$ is a simple $r$-Pareto process with tail index $1/\alpha$ and spectral measure $\sigma_r$, which satisfies the following conditions:
\begin{enumerate}[(1)]
	\item The random variable $r(Z)$ has an $\alpha$-Pareto distribution;
	\item  The random variable $r(Z)$ (spatial aggregate) and the $r$-normalized process $Z/r(Z)$ (spatial profile) are stochastically independent, and $Z/r(Z)$ has probability distribution $\sigma_r$ with support domain $\{f \in \calC_+(\calS): r(f) = 1\}$.
\end{enumerate}
\end{theorem}

Here, we set $\alpha=1$ without loss of generality, which leads to the unit Pareto distribution for margins and the aggregated functional $r(Y)$.  
Note that this choice is not a restriction in practice, as we can always use the estimated marginal model to standardize the data to a common scale, e.g., unit Pareto, using the probability integral transform. 

To be more specific, we here consider a flexible class of $r$-Pareto processes based on log-Gaussian processes. Given a zero-mean Gaussian process $\tilde{X}$ with stationary increments, we write $X(\bm s)=\exp\{\tilde{X}(\bm s)-\mathbb{E}[\tilde{X}(\bm s)^2]/2\}$, which ensures $\mathbb{E}[X(\bm s)]=1$. Based on the process $X$,  we can construct an $r$-Pareto process whose distribution is fully characterized by the combination of $r$, $\alpha$ and $\gamma(\bm h)=\mathbb{E}[\{\tilde{X}(\bm h)-\tilde{X}(\bm 0)\}^2]/2$, the semivariogram of $\tilde{X}$, where $\bm h$ denotes the spatial lag. A general characterization of $r$-Pareto processes for arbitrary risk functionals $r$, and also for the corresponding limits leading to max-stable processes or Poisson point processes, is based on the so-called exponent measure $\Lambda$ defined for Borel sets in $\calC_+(\calS)\setminus \{0\}$, where $0$ refers to the function that is constant and identically equal to zero. Then, the probability distribution of the $r$-Pareto process $Z$ can be written as follows:
$$
\Pr(Z\in (\cdot)) = {\Lambda((\cdot) \cap \{f\in \calC_+(\calS): r(f)\geq 1\})\over\Lambda\{f\in \calC_+(\calS): r(f)\geq 1\}},
$$
which means that the probability measure is obtained by scaling the measure of $\Lambda$ for $r$-exceedances above $1$ (which is known to be positive for $r$-functionals that we use here) to be equal to unity, and by truncating the measure where $r$ does not exceed the threshold $1$. 
For the log-Gaussian construction with $X(\bm s)$ defined as above, the measure $\Lambda$ is fully characterized by the following property, which must hold for any positive function $u(\bm s)\in C_+(\calS)$:
\begin{equation}\label{eq:exponent-measure}
\Lambda\left\{f\in C_+(\calS) : \max_{\bm s\in\calS} {f(\bm s)\over u(\bm s)}>1\right\} = \int_0^\infty \left(1-\Pr\left( \max_{\bm s\in\calS} {X(\bm s)\over u(\bm s)} \leq {1\over r}\right)\right){1\over r^2}\mathrm{d}r.
\end{equation}
By applying the transformation of variable $v=1/r$ and using the fact that $\mathbb{E}[Q]=\int_0^\infty \Pr(Q\geq q)\mathrm{dq}$ for a generic nonnegative random variable $Q\geq0$, the expression in \eqref{eq:exponent-measure} can further be simplified as $\mathbb{E}[\max_{\bm s\in\calS} X(\bm s)/u(\bm s)]$. The probability density of $\Lambda$ when $\calS$ is a finite set of locations is available in closed form for these log-Gaussian extremal models \citep{Engelke.etal:2014}.

\begin{figure*}[!t]
	\includegraphics[width=1\textwidth]{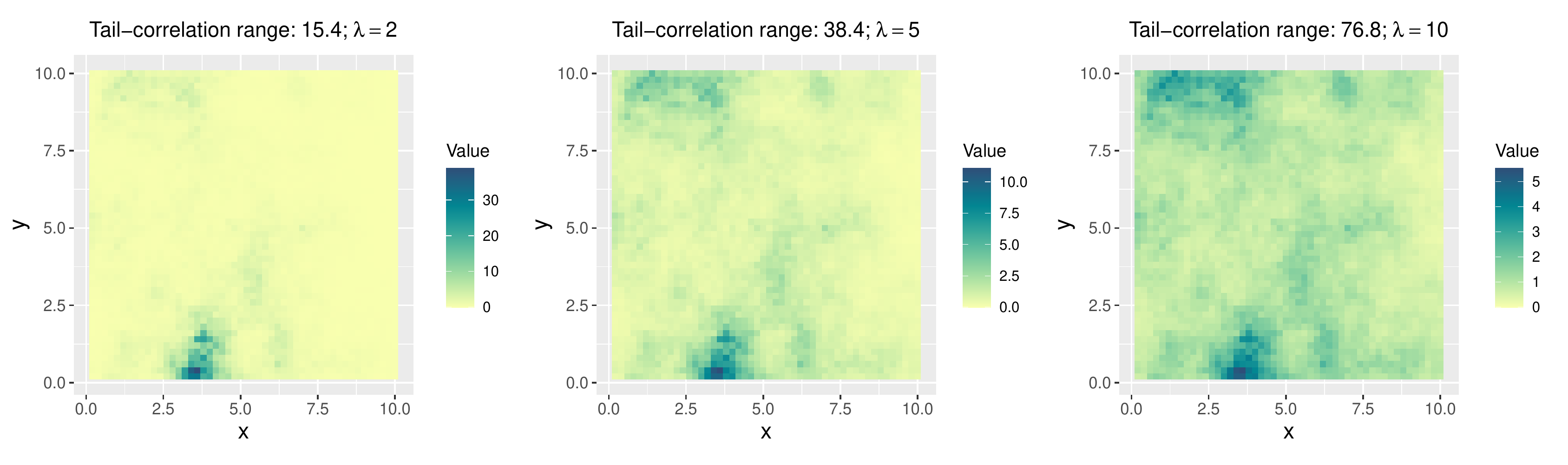}
	\caption{Three simulated $r$-Pareto processes (with the same random seed) on a $50\times50$ grid based on log-Gaussian processes with $\gamma(\bm h) = \|\bm h\|/\lambda$, $\lambda=2,5,10$ (left to right), and $\alpha=5$.}
	\label{fig:simu_dep_range}
\end{figure*}

In this work, our goal is to assess whether the spatial extent of extreme precipitation events has changed over time, and to provide a robust modeling framework that can provide reliable future projections. Therefore, we develop and fit $r$-Pareto processes that can capture non-stationary temporal variations in the spatial dependence structure of extreme precipitation, which directly controls the spatial extent of extreme events. Specifically, we link the spatial dependence range of precipitation extremes on day $t$ to an informative temporal covariate, which is taken to be the same spatiotemporal temperature average, $\text{temp}_t$, proposed and justified in Section~\ref{sec:covariate}. 
Mathematically, we define the (spatial) semivariogram $\gamma$ on day $t$ as 
\begin{equation}\label{eq:semivariogram}
\gamma(\bm h;t) = \left\{{\|\bm h\|\over\exp(\lambda_0 + \lambda_1\times \text{temp}_t)}\right\}^\nu,	
\end{equation} 
where $\bm h$ is the spatial lag vector, $\nu\in(0,2]$ is a smoothness parameter, $\lambda_0\in\Real$ is the baseline spatial range parameter (on the log scale), and $\lambda_1\in\Real$ controls the extent to which the temperature covariate $\text{temp}_t$ affects the range parameter, and thus how the spatial extent of extreme precipitation changes over time. The Gaussian process $\tilde{X}$ in Section~\ref{subsec:dep_model} associated with this semivariogram is known as fractional Brownian motion, for which we can set $\tilde{X}(\bm 0)=0$, such that $\mathbb{E}[\tilde{X}(\bm 0)^2]=0$ in the construction of the process $X$  in  Section~\ref{subsec:dep_model}. 
The spatial extent of extreme precipitation for each time point is measured by the effective tail-correlation range, defined as the minimum distance $\|\bm h\|$ in kilometers such that the tail-correlation coefficient drops below 0.05, i.e., 
\begin{equation}\label{eq:extremal_dep}
	\Pr[\{\hat Y_{\bm s + \bm h,t} \geq u \mid  \hat Y_{\bm s,t} \geq u, r(\{\hat Y_{\bm s,t}\}_{\bm s\in\calS}) \geq u] = 2-2\Phi\left[\left\{{\Gamma(\bm h;t)\over 2}\right\}^{1/2}\right] = 0.05,
\end{equation}
where $\Phi$ is the standard Gaussian cumulative distribution function. Our definition in \eqref{eq:extremal_dep} is analogous to the common ``effective correlation range'' in classical geostatistics, but adapted to extremes based on the well-known tail-correlation $\chi$-measure \citep{Huser.Wadsworth:2022}. To illustrate this concept, Figure~\ref{fig:simu_dep_range} shows three simulated $r$-Pareto processes based on log-Gaussian processes with $\alpha=5$ (i.e., marginal tail index $1/\alpha=0.2$) and variogram $\gamma(\bm h) = \|\bm h\|/\lambda$, on a $50\times 50$ grid, where the different panels display realizations for $\lambda = 2,5,10$ (left to right). We report the associated effective tail-correlation range in the title of each panel. When $\lambda$ increases, the spatial variability of the values clearly decreases and therefore the dependence strength increases, which is also indicated by the values of the effective tail-correlation range.   

Efficient inference for $r$-Pareto processes in high dimensions has been developed by \cite{deFondeville.Davison:2018} using gradient-score matching, a technique that has been implemented in the \texttt{R} package \texttt{mvPot} for stationary $r$-Pareto processes associated with extremal-$t$ max-stable processes and Brown--Resnick max-stable processes. Here, we generalize and extend the \texttt{R} code by \cite{deFondeville.Davison:2018} to incorporate temporal non-stationarity and use it to detect potential temporal trends in the dependence structure. 

\section{Estimation results and future projections}\label{sec:results}
\subsection{Marginal estimation}\label{subsec:marginalestimation}

\begin{figure}[!t]
	\centering
	\includegraphics[width=0.46\linewidth]{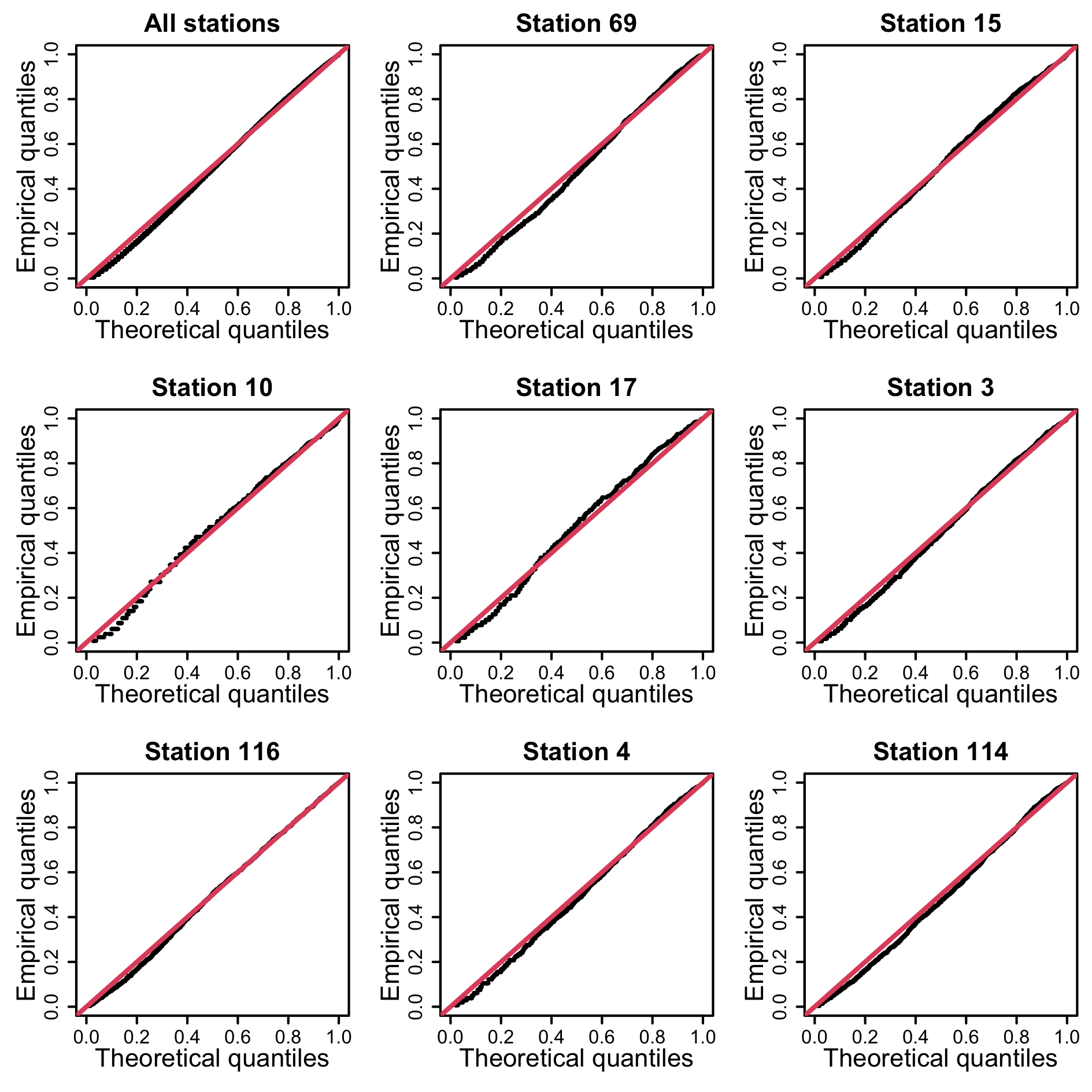}
	\hfill
	\includegraphics[width=0.46\linewidth]{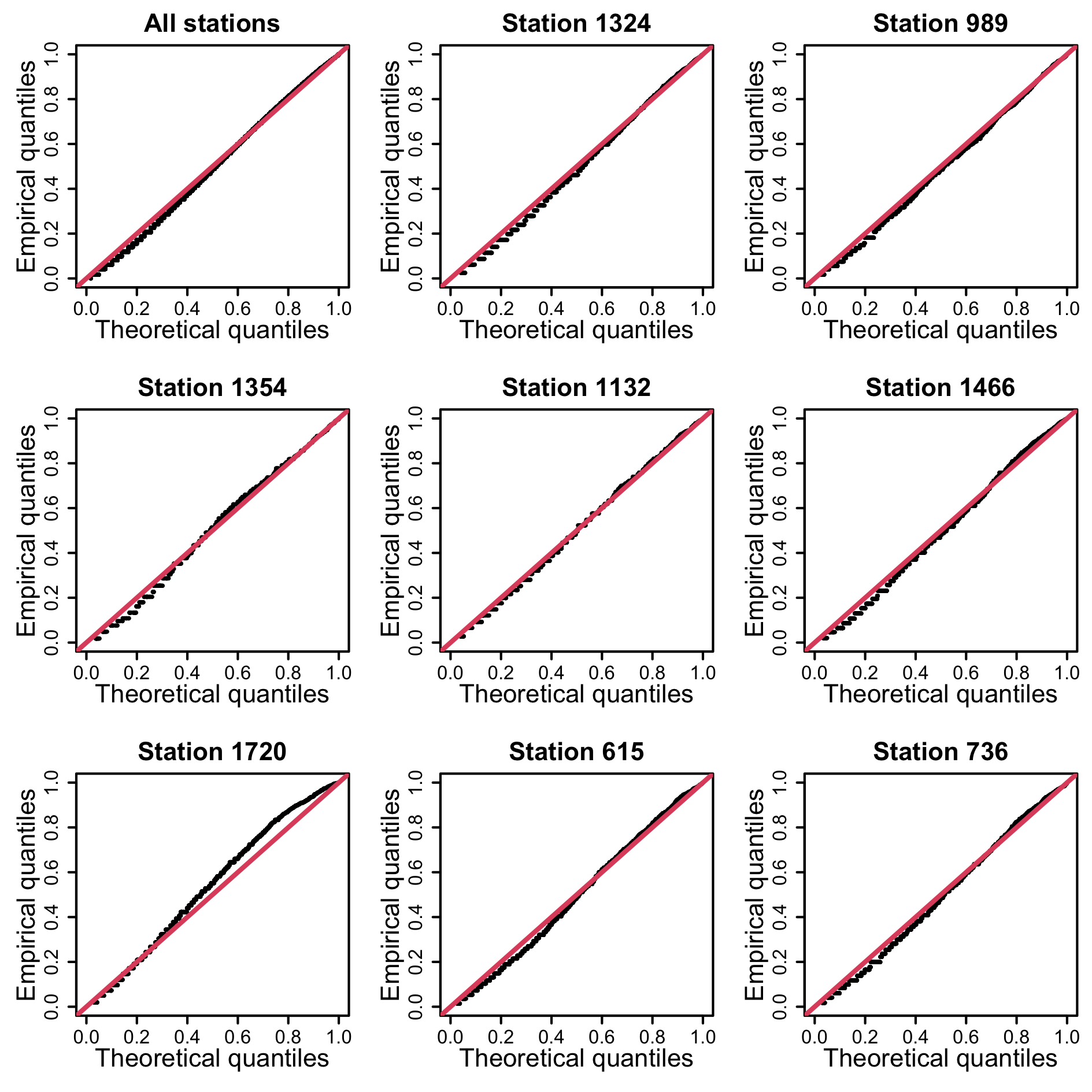}
	\caption{QQ-plots for the marginal fit in the Danube river basin (3 leftmost columns) and the Mississippi river basin (3 rightmost columns).  The data have been transformed to the uniform scale. For each basin (with its corresponding $3\times3$ panels), the top-left QQ-plot shows the data pooled from all the stations in each region. The other eight QQ-plots show the data at randomly selected stations in each basin. }
	\label{fig:marginal_qq}
\end{figure}

To analyze our precipitation data, we start by fitting the three-step model detailed in Section~\ref{subsec:marginal_model} for each of the two river basins considered in this work. After model estimation, we can transform the data to a common scale by exploiting the probability integral transform. To demonstrate the goodness-of-fit of our marginal model, we show Quantile-Quantile (QQ) plots in Figure~\ref{fig:marginal_qq} on the uniform scale. These QQ-plots show that the marginal model fits the data very well overall as the black dots are well aligned along the diagonal. To proceed with the dependence fit (Section~\ref{subsec:dep_model}), we further transform the data to the unit Pareto scale, i.e., with distribution $\Pr(Y \leq y) =1- y^{-1}, y \geq 1$.

\begin{figure}[!t]
	\centering
	\includegraphics[width=0.49\linewidth,page=4]{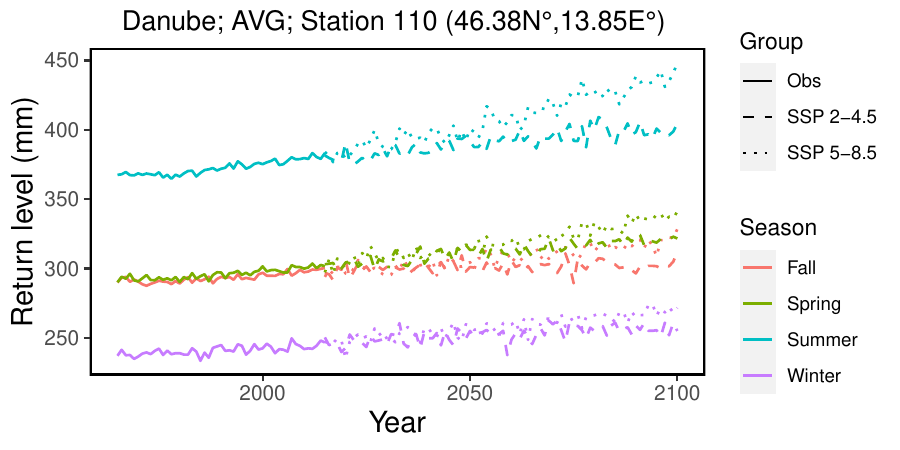}
	\includegraphics[width=0.49\linewidth,page=8]{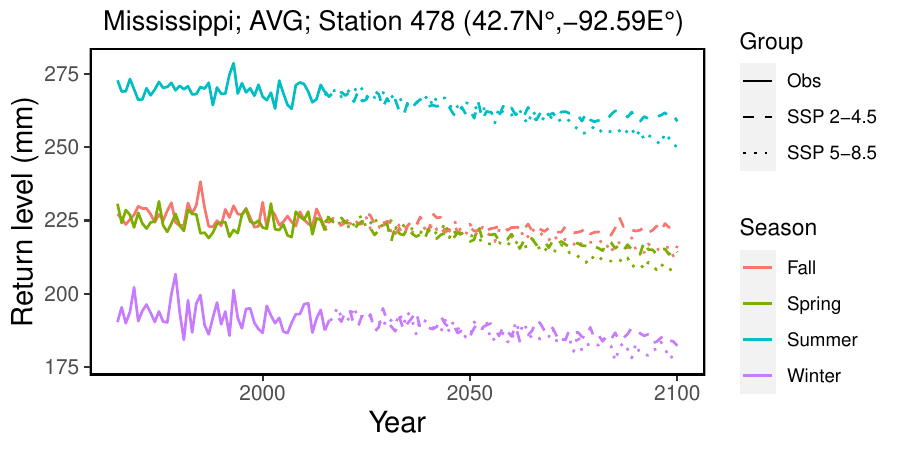}
	\caption{The estimated seasonal average (non-stationary) return level for the exceedance probability $q = 1/(100\times365)$ for each season at Station $110$ in the Danube basin (left), and Station $478$ in the Mississippi basin (right), based on the temperature covariate derived from observational data (solid; 1965--2015) or predicted (2016--2100) from CMIP6 climate model simulations under SSP 2-4.5 (dashed) and SSP 5-8.5 (dotted).}
	\label{fig:marginal_return_level}
\end{figure}

Marginal return levels are commonly used as simple and intuitive measures of marginal risk. The $M$-year return level is defined as the level that is expected to be exceeded once every $M$ years, under stationary conditions. It is simply a high marginal quantile. By analogy, in a changing climate, we can similarly define return levels as time-varying marginal quantiles corresponding to low exceedance probabilities $q$, though the original interpretation is now slightly different. Based on the marginal model from Section~\ref{subsec:marginal_model}, return levels for an exceedance probability $q < 0.1$, denoted by $y_{\bm s,t}^q$, can be estimated for each station $\bm s\in\calS$ and time $t$ as
$$\widehat{y_{\bm s,t}^q} = 10 + \hat u_{s,t} + {\rm GP}^{-1}\left( 1-{q \over p_{\hat u_{s,t}}}; \hat \sigma_{s,t} ,\hat\xi \right),$$
where $\hat u_{s,t}$ is the estimated $90\%$ quantile from the Gamma model, $p_{\hat u_{s,t}}$ is the estimated threshold exceedance probability from the logistic model, and  $\hat \sigma_{s,t}$ and $\hat \xi$ are the estimated scale and shape parameters from the GP model, respectively, with $\text{GP}^{-1}$ indicating the GP quantile function. In both the Danube and Mississippi river basins, the shape parameter estimate is $\hat\xi = 0.12$, which implies that the precipitation distribution is moderately heavy-tailed. Figure~\ref{fig:marginal_return_level} shows the estimated (non-stationary) seasonal averages of return levels for $q = 1/(100\times365)$ (i.e., for a return period of 100 years under stationary conditions) for the stations with the largest average precipitation in the two river basins, namely Station $110$ in the Danube basin and Station $478$ in the Mississippi basin, based on the temperature covariate derived from observational data (1965--2015) or predicted (2016--2100) from CMIP6 climate model simulations under two Social Socioeconomic Pathways (and averaged across three runs). The estimated return levels derived from each individual climate model run are presented in Figure~2 of the Supplementary Material.
Results show a clear seasonality and a discernible increasing trend in the precipitation return levels for the Danube region, especially during the summer time. However, the return levels seem to remain more stable for the Mississippi region, though with a slight negative trend overall. This result is consistent with the results obtained by \cite{Helga.Holger.David:2021}, who have shown that the intensity distribution remains relatively stable for the northeastern United States.  

\subsection{Spatial extreme dependence estimation}\label{subsec:dependenceestimation}
To study seasonal differences in spatial dependence characteristics of extreme precipitation, we proceed by fitting the proposed time-varying $r$-Pareto process separately for each season and each basin. Let $\hat Y_{\bm s,t}$ denote the random precipitation process at time point $t$ and location $\bm s\in \calS$, transformed to the unit Pareto scale by applying the probability integral transform combined with the fitted marginal distribution; recall Sections~\ref{subsec:marginal_model} and \ref{subsec:marginalestimation}. 
As mentioned in Section~\ref{subsec:dep_model},  
extreme precipitation events are defined as $\{ \{\hat Y_{\bm s,t}\}_{\bm s\in\calS}: r(\{\hat Y_{\bm s,t}\}_{\bm s\in \calS}) \geq u\}$, for some high threshold $u$ and risk functional $r$. We here choose the parametric family $r_\theta(\{\hat Y_{\bm s,t}\}_{\bm s\in \calS}) = \left(K^{-1} \sum_{k=1}^K \hat Y_{\bm s_k,t}^\theta\right)^{1/\theta}$, with $\theta>0$, where $\calD=\{\bm s_1,\ldots,\bm s_K\}\subset\calS$ is the set of monitoring stations. Note that $r_\theta$ is a norm only when $\theta\geq1$. We here set either $\theta=1$, which corresponds to summing observations on the Pareto scale across all stations, or $\theta=\widehat\xi\approx0.12$, which first transforms observations back to their original data scale before summing, thereby giving more weight to smaller values. Moreover, for reasons of estimation stability, we only consider spatial replicates $\{\hat Y_{\bm s,t}\}_{\bm s\in \calS}$ that have at least 5 non-missing values at the sites in $\calD$. 
The threshold $u$ is selected to be the $80\%$ empirical quantile of observed $r_\theta(\{\hat Y_{\bm s,t}\}_{\bm s\in \calS})$ values, computed from at least 5 non-missing values across stations. As we have already discarded precipitation amounts that are smaller than 10 mm in the marginal modeling phase, the $80\%$ quantile is relatively extreme and still provides sufficient data points to estimate the spatial tail dependence structure. The resulting numbers of selected events in each basin--season case are summarized in Table~\ref{tab:number_events}. Note that while the number of selected events is the same for both $r$-functionals, namely $r_1$ and $r_{\widehat\xi}$, the selected events themselves are not identical, and the number extreme events common to both risk functionals is presented in parentheses. About half of those selected events are the same using these two risk functionals.   

\begin{table}[!t]
\caption{Number of spatial extreme events selected for each basin and season. Note that the number of selected events is the same for both risk functionals, $r_1$ and $r_{\widehat\xi}$, and the number of extreme events that are common to both risk functionals is shown in parentheses.}
\centering
\begin{tabular}{lcccc}
\toprule
\textbf{Basin} &  \textbf{Winter} & \textbf{Spring} & \textbf{Summer} & \textbf{Fall} \\
\midrule
\textbf{Danube} & 62 (40) & 118 (76)  & 133 (90) &  100 (59)  \\
\textbf{Mississippi} & 115 (67) & 225 (97) & 232 (118) & 206 (98)	\\
\bottomrule
\end{tabular}
\label{tab:number_events}
\end{table}

\begin{table}[t] 
\centering
\caption{Parameter estimates $\widehat\nu,\widehat\lambda_0,\widehat\lambda_1$ of the $r$-Pareto model \eqref{eq:semivariogram} fitted to spatial extreme events from each season and river basin, as well as for both risk functionals, $r_1$ and $r_{\widehat\xi}$. The numbers within parentheses are $95\%$ confidence intervals based on 300 nonparametric bootstrap fits and significant estimates of the trend coefficients $\lambda_1$ are highlighted in bold.}
\label{tab:estimates}
\begin{adjustbox}{width=1\linewidth,center}
\begin{tabular}{lccccccc}
\toprule
\textbf{Basins} &  & \multicolumn{3}{c}{\textbf{Danube}} & \multicolumn{3}{c}{\textbf{Mississippi}} \\ 
\cmidrule{3-5} \cmidrule{6-8} \\
& $r$ & $\hat\nu$ & $\hat\lambda_0$ & $\hat\lambda_1$ & $\hat\nu$ & $\hat\lambda_0$ & $\hat\lambda_1$ \\
\midrule
\textbf{Winter} & $r_1$ & 0.29 & 2.60 & 0.16 & 0.29 & 3.97 & 0.02 \\
&  & (0.24,0.36) & (2.00,3.10) & (-0.21,0.51) & (0.26,0.32) & (3.75,4.20) & (-0.19,0.26) \\
& $r_{\hat\xi}$ & 0.24 & 3.74 & 0.01 & 0.26 & 5.48 & 0.11 \\
&  & (0.20,0.29) & (3.37,4.06) & (-0.34,0.40) & (0.24,0.28) & (5.29,5.67) & (-0.06,0.31) \\
\midrule
\textbf{Spring} & $r_1$ & 0.26 & 2.17 & -0.24 & 0.28 & 4.07 & \textbf{-0.17} \\
& & (0.21,0.34) & (1.61,2.71) & (-0.62,0.09) &(0.26,0.30) & (3.93,4.20) & (-0.29,-0.05)  \\
& $r_{\hat\xi}$ & 0.23 & 3.61 & \textbf{-0.34} & 0.27 & 4.98 & \textbf{-0.49}  \\
& & (0.19,0.29) & (3.25,3.89) & (-0.67,-0.06) & (0.26,0.28) & (4.88,5.07) & (-0.59,-0.38) \\
\midrule  
\textbf{Summer} & $r_1$ & 0.24 & 1.81 & -0.23 & 0.24 & 3.22 & -0.07 \\
& & (0.18,0.34) & (0.93,2.62) & (-0.70,0.21) & (0.22,0.26) & (3.05,3.40) & (-0.22,0.06) \\
& $r_{\hat\xi}$ & 0.23 & 3.57 & -0.23 & 0.24 & 4.12 & \textbf{-0.18} \\
& & (0.19,0.33) & (3.20,3.90) & (-0.55,0.09) & (0.23,0.25) & (4.04,4.21) & (-0.30,-0.09) \\
\midrule
\textbf{Fall} & $r_1$ & 0.27 & 2.78 & 0.22 & 0.27 & 4.32 & 0.00 \\
& & (0.22,0.44) & (2.19,3.37) & (-0.18,0.56) & (0.26,0.29) & (4.12,4.51) & (-0.20,0.20) \\
& $r_{\hat\xi}$ & 0.27 & 3.99 & 0.25 & 0.28 & 5.52 & -0.09 \\
& & (0.23,0.34) & (3.67,4.23) & (-0.06,0.49) & (0.27,0.29) & (5.37,5.66) & (-0.22,0.05) \\
\bottomrule
\end{tabular}
\end{adjustbox}
\end{table}

\begin{figure}[!t]
 \includegraphics[width=1\linewidth]{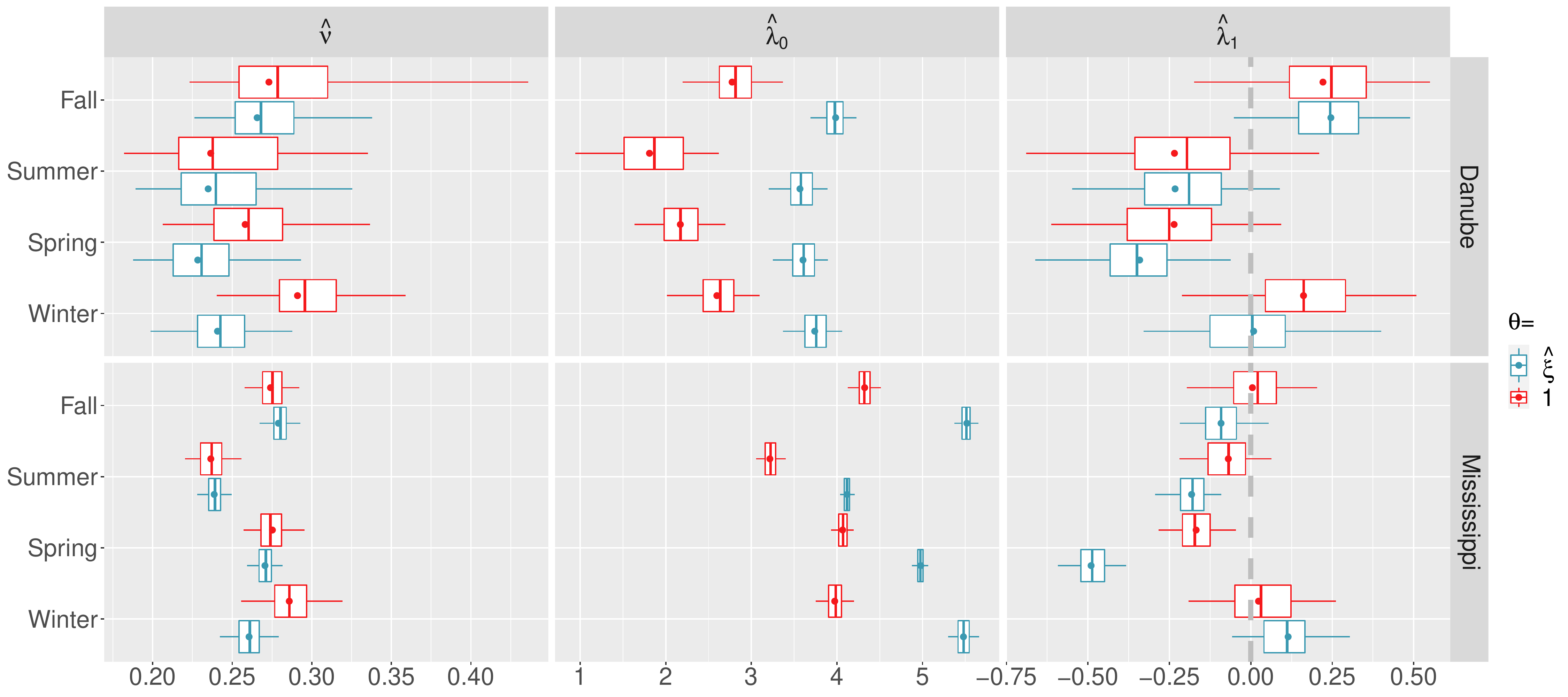}
\caption{Parameter estimates $\widehat\nu,\widehat\lambda_0,\widehat\lambda_1$ (left to right) from the $r$-Pareto dependence model \eqref{eq:semivariogram} fitted to spatial extreme events from the Danube (top) and Mississippi (bottom) basins for each season, based on the risk functionals $r_1$ (red) and $r_{\widehat\xi}$ (blue). Each panel displays modified boxplots of the 300 nonparametric boostrap estimates showing $2.5\%$, $25\%$, $50\%$, $75\%$ and $97.5\%$ quantiles, as well as point estimates from the original data (solid dots). The dashed grey vertical line at zero in the $\hat\lambda_1$ plots represents the ``no trend'' reference.}
\label{fig:CIPlot}
\end{figure}

Table~\ref{tab:estimates} and Figure~\ref{fig:CIPlot} report parameter estimates from the $r$-Pareto process fit for each basin and season, as well as both $r$-functionals for comparison. Bootstrap-based $95\%$ confidence intervals are shown in parentheses in Table~\ref{tab:estimates} and illustrated graphically through modified boxplots in Figure~\ref{fig:CIPlot}. More precisely, these confidence intervals are computed from 300 nonparametric bootstrap fits, whereby the spatial extreme events are resampled with replacement and the $r$-Pareto dependence model refitted to the resampled extreme events 300 times. The modified boxplots here display five empirical quantile levels (namely $2.5\%$, $25\%$, $50\%$, $75\%$, and $97.5\%$) from the nonparametric bootstrap estimates, while the solid dots in each modified boxplot represent point estimates based on original data. Interestingly, we find that the estimated regression coefficient $\hat\lambda_1$ (i.e., the slope of the temperature covariate, $\text{temp}_t$) is always negative for both the Danube and Mississippi basins during the major rain seasons, i.e., summer and spring, especially when $\theta = \widehat\xi$. In this case, the $\hat\lambda_1$ estimates are statistically significant in three out of the four cases (Danube in Spring, Mississippi in Spring, and Mississippi in Summer) and the last case (Danube in Summer) misses significance by a small margin only. Therefore, our results indicate that the spatial dependence range (and thus, the spatial extent of extreme precipitation) tends to decrease as temperature increases during the major rain seasons. During winter and fall, our results suggest that an opposite pattern usually tends to prevail for the two basins (i.e., with slightly positive regression coefficients), though the effects are not statistically significant. We also observe that, overall, our results are consistent across the risk functionals $r_1$ and $r_{\widehat\xi}$, regardless of the statistical significance of the $\hat\lambda_1$ estimates, which suggests that our conclusions are robust to the definition of ``spatial extreme event''.  
  
In summary, our findings show decreases in precipitation extents with increases in temperature in both the Mississippi region and the Danube region during the major rain seasons, which suggests a further decline in a warming climate. 
In the next section, we further exploit our fitted model to investigate potential changes in spatial extents of extreme precipitation. Specifically, we make projections until the end of the 21st century by relying on our fitted $r$-Pareto process combined with the proposed temperature covariate stemming from (debiased) climate model outputs under different climate change scenarios.

\subsection{Future spatial extent projections}\label{subsec:futureprojections}
To evaluate the spatial extent of extreme precipitation in the future, we here construct a temperature covariate similar to the one used for training the model by exploiting daily temperatures predicted by climate models from the sixth Coupled Model Intercomparison Project (CMIP6). We considered a subset of global circulation models (GCMs) following suggestions by \cite{Brunner.Pendergrass.etal:2020} who provide a ranking of GCMs based on historical performance, and eventually selected three GCMs that are highly ranked and also available in the Copernicus data base (\url{https://www.copernicus.eu/en/access-data}) with relatively high resolution. Specifically, the selected GCMs are AWI-CM-1-1-MR (AWI), MIROC6 (MIROC), and NorESM2-MM (NorESM) with resolutions of $0.9351^\circ\times0.9375^\circ$, $1.4008^\circ\times1.4072^\circ$, and $0.9424^\circ\times1.25^\circ$ approximately in latitude and longitude, respectively. We also consider two future climate change scenarios, known as Shared Socioeconomic Pathways (SSPs), and specifically choose relatively optimistic (SSP2-4.5) and pessimistic (SSP5-8.5) scenarios. To be consistent with the observed temperature covariate used to fit the models, we then apply the same kriging scheme as described in Section~\ref{sec:covariate} to the climate outputs. After kriging, the corresponding spatiotemporal temperature values are averaged and renormalized as described in Section~\ref{sec:covariate}. Since simulated temperature from GCM outputs is often subject to systematic biases, and thus not perfectly aligned with observed temperatures, it is necessary to bias-correct climate model outputs. Here, we adopt a basic bias-correction approach and simply subtract, for each season and basin separately, the difference between the average temperature of GCM simulations over 2015--2020 and observations over 2010--2015 by assuming that there is no significant change in the temperature averages over the periods 2010--2015 (for observations) and 2015--2020 (for simulations). While this approach is quite basic, it adequately removes the unrealistic visual discontinuity between observations and simulations in 2015, while preserving the overall temporal trend. This procedure yields six possible temperature covariates (i.e., one per GCM type and SSP scenario) for each season and basin, that we can exploit for future extrapolation. To summarize that information concisely and to derive a ``representative'' temperature covariate across the GCMs considered, we also compute the average projected temperature across the three GCMs, but separately for each SSP scenario. We here only report the results from the average across GCMs, and present the separate GCM-specific results in the Supplementary Material. 

\begin{figure}[!t]
	\centering
	\includegraphics[width=0.7\linewidth,page=4]{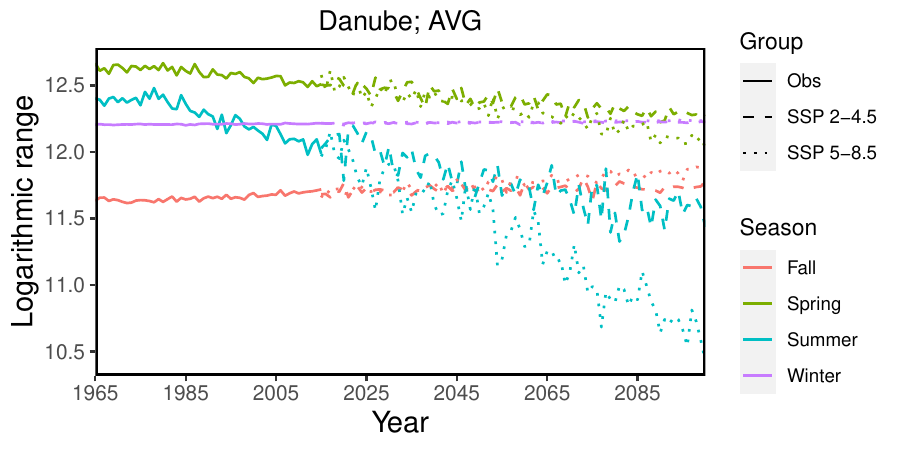}
	\includegraphics[width=0.7\linewidth,page=8]{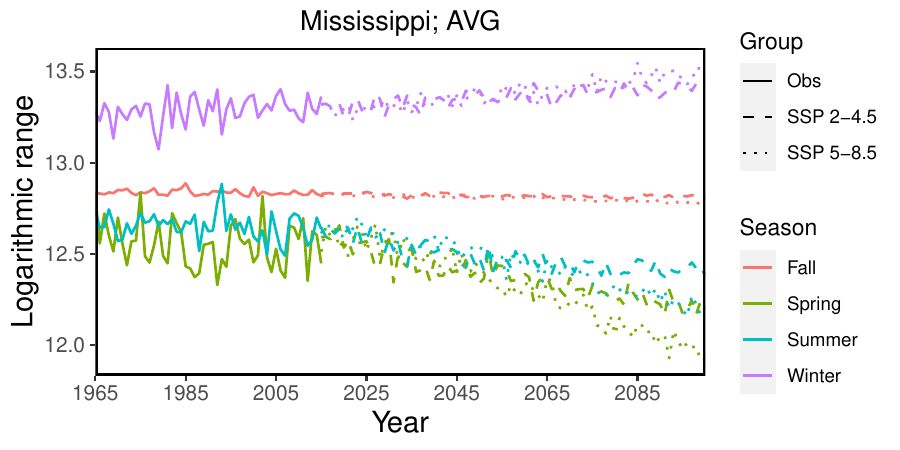}
 \caption{Effective tail-correlation range (km) in logarithmic scale for the average (AVG) of three climate model outputs, AWI, MIROC, and NorESM in the Danube region (top) and the Mississippi region (bottom) for each season (Winter, Spring, Summer, Fall) based on the observed temperature averages (square), optimistic climate change scenario (SSP2-4.5, dashed), and pessimistic climate change scenario (SSP5-8.5, dotted).}
	\label{fig:extreme_range} 
\end{figure}

Using our debiased climate model-based temperature covariate, we compute the effective tail-correlation range for each time point, defined earlier in \eqref{eq:extremal_dep}.
Figure~\ref{fig:extreme_range} shows the projected effective tail-correlation range on the logarithmic scale, estimated from our fitted model based on the $r_{\widehat\xi}$-functional, for each season, each basin, and the two different climate change scenarios (SSP2-4.5 and SSP5-8.5) after averaging the temperature covariate across GCMs (recall Figure~\ref{fig:temp_covariate}). In the Supplementary Material, we show figures of the projected effective tail-correlation range that is derived from each individual GCM. The plots show that differences between the changes in the effective tail-correlation range under the SSP 2-4.5 and SSP 5-8.5 scenarios are larger for the Danube basin than for the Mississippi basin. These findings suggest that climate change may have stronger impacts on spatial precipitation extents in the Danube region than in the Mississippi region. As suggested by the $\hat\lambda_1$ estimates from Table~\ref{tab:estimates}, Figure~\ref{fig:extreme_range} shows that the effective tail-correlation range is expected to decrease in a warming climate during the major rain seasons, both in the Danube and Mississippi regions, with more dramatic changes under the SSP5-8.5 scenario.

\section{Discussion}\label{sec:discussion}
\subsection{Statistical considerations}
Statistical modeling of spatial extremes is a complex task, and it becomes even more complex in the setting of temporal nonstationarity due to the different possibilities to model nonstationarity and the required higher model sophistication that follows from it. We have here focused on using a  covariate with strong physical motivation,  and we have embedded it into model parameters in a linear way (modulo the use of link functions). Therefore, the form of nonstationarity captured by our model is restricted to the chosen temporal covariate, and it cannot identify nonlinear effects that are more complex than the covariate itself. However, our approach can be considered as robust for estimation and extrapolation beyond the range of covariate values used for training. 

The calculation of a basin average for the temperature covariate allowed us to use different types of temperature data for the training and projection period, respectively. Locally, climate model outputs are usually ``smoother" than actual observations at weather stations, but the averaging step led to a comparable degree of smoothing across the whole basin for both data types. It would be interesting to refine our approach by using local temperature covariates that are not averaged over the whole basin, such that intra-basin differences in temperatures would be better taken into account and could lead to refined local interpretations of results. However, to avoid biases with our approach, this would require having realistic climate model outputs at relatively small spatial resolution of at most several kilometers, which is not yet the case for most climate models of the current generation. Regional downscaling approaches, such as those of the recent CORDEX initiative \citep{Giorgi:2015}, already enable working at relatively smaller scales of around $10$km and convection-permitting models provide data at event finer spatial resolutions of 2--3km \citep{Lucas.etal:2021,Coppola.etal:2020}. 

Handling intermittence with absence of precipitation at some stations, and handling low precipitation values that are observed imprecisely (e.g., with discretization effects due to rounding of values),  is notoriously difficult in statistical modeling and estimation. We have here opted for a relatively simple procedure by removing precipitation observations during extreme episodes if they are below 10mm to improve marginal fits. Since this approach flags observations as missing while there actually is information about them being very small, our models might tend to slightly overestimate the spatial dependence range of extremes. However, given that we treat all basins, seasons and time periods in the same way, we can expect similar biases in all cases, such that the results, and especially the interpretation of estimated dependence ranges across basins, seasons and time periods, should not be affected. 

With $r$-Pareto processes, we have used models for extremal dependence that arise asymptotically in extreme-value theory and therefore provide a sound and robust modeling framework in the extreme-value setting where data are not abundant. A key feature of such models is that the spatial extent of extreme episodes remains constant (on average) when moving towards higher quantiles of the risk functional $r$ (with all model parameters being held fixed). Many recent works on modeling environmental extremes rather suggest that spatial extents often tend to decrease as the level of risk functionals increases \citep[e.g.,][]{Wadsworth.Tawn:2012,Opitz:2016,Huser.etal:2017,Tawn-et-al-2018,Bacro.al.2020,Huser.etal:2021,Huser.Wadsworth:2022,Zhang.etal:2022}. However, here we do not use our models to extrapolate far into the tail, but we rather study the spatial characteristics of extremal dependence at high but fixed and finite quantile levels. Hence, this potential model misspecification with respect to asymptotic dependence stability is not problematic for our approach and for the insights we gain from it. Future extensions of the current work could consider the use of more flexible ``subasymptotic'' models reviewed by \citet{Huser.Wadsworth:2022}; however, the appealing use of general risk functionals to target the modeling of specific types of extreme-event episodes is not yet possible with most of these approaches.

\subsection{Hydro-meteorological considerations}
While our findings show an increase in local precipitation intensities with increases in temperature, they show a decrease in precipitation dependence during the main rain season. As a consequence, further decreases of precipitation extents are projected in a warming climate. Furthermore, our analysis for two river basins in different hydro-climates and different seasons highlights that the relationship between temperature and precipitation extent is to some degree season- and region-dependent. For example, relationships between temperature and precipitation extent are more pronounced in the Mississippi than in the Danube river basin and in spring and summer compared to winter and fall.
These spatiotemporal variations in the relationship between temperature and spatial precipitation extent may be the reason for the disagreement of past studies on the direction of change in spatial precipitation extent with increasing temperatures.
While some observation- and simulation-based studies have shown an increase of spatial precipitation extent over time for some seasons and regions \citep{Bevacqua.etal:2021,Tan.etal:2021, Rastogi.etal:2020, Lochbihler.etal:2017, Dittus.etal:2015, Nikumbh.etal:2019}, others have shown a decrease for other seasons and regions \citep{Benestad:2018, Guinard.etal:2015, Wasko.etal:2016}. In addition, while some of the studies showing increases in precipitation extent focused on the winter season \citep[e.g.,][]{Bevacqua.etal:2021}, some of the studies showing decreases in extent focused on summer \citep[e.g.,][]{Guinard.etal:2015}. Similar seasonal variations have also been found in previous studies assessing the ``length scales'' of extreme precipitation \citep{Touma.etal:2018}. Such seasonal variations in the changes in precipitation extent suggest that changes in precipitation extent may be related to changes in weather patterns and storm types. For example, convective and localized summer storms may become more frequent in a future climate leading to a decrease in spatial precipitation extents in summer, while storm types may increase in size in other seasons \citep{Chang.etal:2016, Moron.etal:2021}. In addition to season and region, other factors may help to explain the divergent results on changes in precipitation extents by different studies, including method choices, model simulations, and storm selection criteria \citep{Rastogi.etal:2020}.

The projected changes in spatial precipitation extents have potential implications for the spatial extent of flooding in a future climate. There is first evidence for past changes in spatial flood extents over Europe \citep{Berghuijs.etal:2019, Kemter.etal:2020}, yet, it is less clear which hydro-climatic variables caused these changes.
Precipitation is one important flood driver, particularly for high-magnitude events \citep{Berghuijs.etal:2019, Brunner.etal:2021}. However, land-surface processes such as soil moisture and snowmelt modulate spatial flood dependencies in addition to precipitation \citep{Brunner.etal:2020, Rupp.etal:2021}. That is, the detected changes in precipitation extents do likely not directly translate to changes in flood extents.
Still, considering these dependencies is crucial to avoid under- or over-estimating the risk of widespread flooding \citep{Thieken.etal:2015, Brunner.etal:2020b}.




\section{Conclusions}\label{sec:conclude}
In this paper, we used $r$-Pareto processes to model extreme precipitation in two major river basins and to study the time evolution of their spatial extent. To do so, we studied the relationship between temperature and the spatial extent of extreme precipitation by linking a suitable temperature covariate to the range parameter in the underlying semivariogram function. 
Our results show a negative correlation between the spatial precipitation extent and the temperature covariate for the two river basins, Danube and Mississippi, during the major rain seasons. As for the marginal fits, the fitted return level in the Danube river basin is slightly increasing, especially during summer. By contrast, return levels in the Mississippi river basin remain relatively stable. That is, while precipitation intensities are increasing locally in the Danube region or remain stable in the Mississippi region, the spatial extent of precipitation is decreasing with increasing temperatures during major rain seasons. As a consequence, climate simulations based on future temperature scenarios project future decreases in spatial precipitation extents as a result of increasing temperatures. These results are to a certain degree region- and season-specific and generalizations to other regions are challenging.
Simultaneous increases in local precipitation intensities and decreases in spatial extent suggest more localized impacts. In future research, a completely different approach based on self-exciting point processes or spatial logistic regression models could be used to directly assess the time-varying occurrence of extreme precipitation events above a high level.

For reproducibility purposes and to make our methodology more accessible to the whole community, we have made our \texttt{R} code and the data available for download from the following GitHub repository: \url{https://github.com/PangChung/SpatialScalePrecipExtremes}.

\section*{Acknowledgments}
The research reported in this publication was supported by funding from King Abdullah University of Science and Technology (KAUST) Office of Sponsored Research (OSR) under Award No. OSR-CRG2020-4394.


\bibliographystyle{CUP}
\baselineskip=12pt
\bibliography{Total}

\end{document}

%% file: macros.tex
\def\frac#1#2{{\textstyle{#1\over#2}}}

\DeclareSymbolFont{AMSb}{U}{msb}{m}{n}
\DeclareMathSymbol{\Natural}{\mathbin}{AMSb}{"4E}
\DeclareMathSymbol{\Integer}{\mathbin}{AMSb}{"5A}
\DeclareMathSymbol{\Real}{\mathbin}{AMSb}{"52}
\DeclareMathSymbol{\Rational}{\mathbin}{AMSb}{"51}
\DeclareMathSymbol{\Imaginary}{\mathbin}{AMSb}{"49}
\DeclareMathSymbol{\Complex}{\mathbin}{AMSb}{"43} 
\DeclareMathSymbol{\Disk}{\mathbin}{AMSb}{"44} 
\def\bi{\begin{itemize}}
\def\ei{\end{itemize}}
\def\bd{\begin{description}}
\def\ed{\end{description}}
\def\ben{\begin{enumerate}}
\def\een{\end{enumerate}}
\def\bv{\begin{verbatim}}
\def\ev{\end{verbatim}}

\def\calC{{\mathcal C}}
\def\calD{{\mathcal D}}

\def\calS{{\mathcal{S}}}

\def\bar#1{{\overline{#1}}}
\def\hat#1{{\widehat{#1}}}

\def\pr{{\rm Pr}}
\def\Pr{\pr}

\def\2to{{\ {\buildrel 2\over \longrightarrow}\ }}

\def\I1ton{{$I_1,\ldots,I_n$}}
\def\X1ton{{$X_1,\ldots,X_n$}}
\def\Y1ton{{$Y_1,\ldots,Y_n$}}
\def\Z1ton{{$Z_1,\ldots,Z_n$}}
\def\R1ton{{$R_1,\ldots,R_n$}}
\def\e1ton{{$e_1,\ldots,e_n$}}
\def\t1ton{{$t_1,\ldots,t_n$}}
\def\x1ton{{$x_1,\ldots,x_n$}}
\def\y1ton{{$y_1,\ldots,y_n$}}
\def\z1ton{{$z_1,\ldots,z_n$}}

\def\calS{{\mathcal{S}}}

%% file: teachingmacros.tex
\newtheorem{defn}{Definition}

\newtheorem{theorem}[defn]{Theorem}